\documentclass[twocolumn,showpacs,aps,epsfig]{revtex4}

\usepackage{graphicx}
\usepackage{epstopdf}
\usepackage{latexsym}

\usepackage[center]{subfigure}

\begin{document}

 \newcommand{\bq}{\begin{equation}}
 \newcommand{\eq}{\end{equation}}
 \newcommand{\bqn}{\begin{eqnarray}}
 \newcommand{\eqn}{\end{eqnarray}}
 \newcommand{\nb}{\nonumber}
 \newcommand{\lb}{\label}
\newcommand{\PRL}{Phys. Rev. Lett.}
\newcommand{\PL}{Phys. Lett.}
\newcommand{\PR}{Phys. Rev.}
\newcommand{\CQG}{Class. Quantum Grav.}




\title{Orbifold branes   in   string/M-Theory and their cosmological applications}

\author{Anzhong Wang}
\email{anzhong_wang@baylor.edu}
\affiliation{  GCAP-CASPER, Physics Department, Baylor University,
Waco, TX 76798-7316 }

\date{\today}
\begin{abstract}

In this brief report, we summarize our recent studies in brane cosmology  in  
both string theory and M-Theory on $S^{1}/Z_{2}$. In such setups, we find
that  the radion is stable and its mass, with a very conservative estimation,  
can be of the order of $0. 1 \sim 0.01$  GeV. The hierarchy problem can be 
addressed by combining the large extra dimension, warped factor, and tension 
coupling mechanisms. Gravity is localized on the visible brane, and the spectrum 
of the gravitational Kaluza-Klein (KK) modes is discrete and can have a mass 
gap of TeV. The corrections to the 4D Newtonian potential from the higher 
order gravitational KK modes are exponentially suppressed. Applying such 
setups to cosmology, we find that a late transient acceleration of the universe 
seems to be the generic feature of the theory, due to the interaction between 
branes and bulk. A bouncing early universe is also rather easily realized.

\end{abstract}

\maketitle


{\em Introduction}:  A   long-standing problem  in physics is  {\em the hierarchy problem}, which has been
one of the main driving forces of physics beyond the Standard Model (SM) during the last few 
decades \cite{Anto07}. The problem can be formulated as the large difference in magnitudes
between the Planck and electroweak scales, ${M_{pl}}/{M_{EW}} \simeq 10^{16}$, 
where $M_{pl} (\sim 10^{16} \; TeV)$ denotes the four-dimensional Planck mass,  
and $M_{EW} ( \sim  TeV)$  the electroweak scale.  

To resolve this problem, in 1998 brane world scenarios were proposed and have been extensively  studied
since then  \cite{branes}. In particular, Arkani-Hamed {\em et al} (ADD) \cite{ADD98} pointed out 
that the extra dimensions need not necessarily be small and may even be on the scale of
millimeters \cite{Boyle}. 
For a typical size $R$ of the extra dimensions, the  D-dimensional fundamental Planck mass 
$M_{D}$ is related to  $M_{pl}$ by 
$M_{D} = \left({M^{2}_{pl}}/{R^{D-4}}\right)^{1/(D-2)}$.
Clearly, for any given   extra dimensions  ($D \ge 6$), if $R$  is large enough, $M_{D}$ can be 
as low as the electroweak scale. 
In a different model, Randall and Sundrum (RS) \cite{RS1} showed that if the self-gravity of 
the brane is included, gravitational effects can be localized near the Planck (invisible) brane at low energy 
and the 4D Newtonian gravity is reproduced. 
 In this model,  often referred to as the 
RS1 model,  the extra dimensions are not homogeneous, but warped.  The mechanism to solve 
the hierarchy problem is  different \cite{RS1}. Instead of using large extra dimensions, RS used 
the warped factor, for which the mass $m_{0}$ measured on the Planck brane is related 
to the mass $m$ measured on the visible (TeV) brane by 
$m = e^{-ky_{c}}m_{0}$, 
where $e^{-ky_{c}}$ is the warped factor. Clearly, by properly choosing the distance $y_{c}$  between the two branes, one 
can lower $m$ to the order of TeV, even $m_{0}$ is still of the order of $M_{pl}$.  A  remarkable 
feature   of the RS1 and ADD models is that, after ten years  of its invention,  no  
evidence of tension with current observations or tests of gravitational phenomena has been found
 \cite{branes}. More important,  they are experimentally testable at high-energy particle colliders and 
 in particular at the newly-built Large  Hadron Collider (LHC)   \cite{DHR00}. 

Another important problem  is {\em the late cosmic acceleration of the universe}, first observed from Type Ia 
supernovae measurements \cite{dark}, and  confirmed subsequently by more detailed studies of 
supernovae  and independent evidence from the cosmic microwave background radiation (CMB)
 and large scale structure. While the  late cosmic acceleration of the universe is now 
well established \cite{WMAP7}, the underlying physics remains a complete mystery \cite{DEs}. The nature and origin of the
acceleration have profound implications, and understanding them is one of the biggest challenges of 
modern cosmology  \cite{DETF}.  

Various models have been proposed \cite{DEs}, which   can be divided into two classes: one is constructed 
within general relativity  (GR), such as quintessence \cite{quit} and a tiny positive cosmological constant (CC),  
and the other is from modified theories of gravity, such as the DGP brane models \cite{DGP}  and   
the $f(R)$ nonlinear gravity \cite{EFP01}.  However,   it is fair to say that so far no  convincing model has been 
constructed, yet. A tiny   CC might be one of the simplest resolutions of the crisis, and is  consistent 
with all observations carried out so far \cite{dark,WMAP7}. It is exactly because of this triumph that,
together with an early inflationary  and subsequently radiation and cold dark matter dominated periods, this model 
has been considered as the current ``standard  model" of cosmology.

Brane world scenarios have been intensively studied in the last decade or so, and 
continuously been one of the most active frontiers of physics. As a matter of fact, the field has been so extensively studied 
that it is very difficult to provide  a list of un-biased references, so we shall simply refer readers to the review articles
\cite{branes}. However,  most of the work carried out so far   is phenomenological in nature. Therefore, it is very important
to consider these models in a ``bigger picture." At the present, string/M theory is our best bet for a consistent  
quantum theory of gravity, so it is  natural to embed such models into  string/M theory. In fact, the invention of branes 
\cite{ADD98,RS1} was originally motivated by string/M theory \cite{HW96}. However,  relatively much less  efforts  have 
been devoted to such studies \cite{LOSW99}. One of the main reasons is that such an embedding  is non-trivial and frequently 
hampered by the complexity of string/M theory. In such a setup, two fundamental  issues are: (a) the stability
of  the large number of moduli resulting from the compactification; and (b)  the localization of  gravity on the  branes. In 
the Horava-Witten (HW) heterotic M theory, the moduli are naturally split into two families, depending on the type of harmonic form, ($1,1$) 
and ($2,1$). Various mechanisms to stabilize these moduli  have been proposed. In particular, one may use the internal fluxes, 
introduced by Kachru {\em et al}  (KKLT) originally for the moduli stabilization of type IIB string \cite{KKLT}, to stabilize the ($2, 1$) 
moduli. Brane stretching between the two boundaries, on the other hand, can fix the ($1,1$) moduli \cite{BO04}, while gaugino 
condensation may fix the volume of the 3 Calabi-Yau manifold \cite{DRSW}. In addition, to fix the distance (radion) between 
the two branes, Goldberger-Wise (GW) mechanism \cite{GW99} and Casimir energy contributions \cite{Toms} have also 
been widely used.  


In the past couple of years, we have   studied orbifold branes in the framework of the 11D  
HW heterotic M Theory on $S^{1}/Z_{2}$ \cite{HW96}, developed by Lukas {\em et al}  \cite{LOSW99} by compactifying the 11D 
HW theory on a 6D Calabi-Yau space \cite{GWW07,WGW08}, and orbifold branes  in the framework of (type II) string theory 
\cite{WS07,WS08,WSVW08,Devin09}, as well as orbifold branes in the RS setup \cite{Wang02,WCS08}. In particular, 
 we
investigated  in detail the three important issues: (i) the radion stability and radion mass; (ii) the localization of 
gravity and high-order Yukawa corrections; and (iii)  the hierarchy problem. 

\begin{figure}[tbp]
\includegraphics[width=\columnwidth]{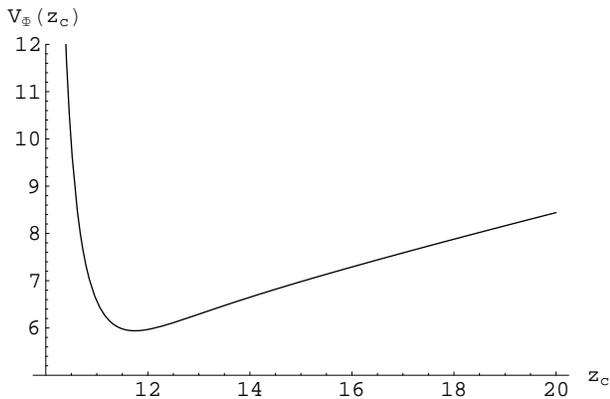}
\caption{The radion potential in the framework of  the HW heterotic M Theory \cite{WGW08}.  }
\label{fig1}
\end{figure}

{\em Radion Stability and masses}: Using the Goldberger and Wise (GW) mechanism \cite{GW99}, we found that the radion is
 stable. Fig. \ref{fig1} shows the radion potential 
in the case of HW heterotic M Theory \cite{WGW08}. A similar result was also obtained in the framework of string theory \cite{WS08}. 

The radion masse is given by \cite{WGW08}
\bq
\lb{2.1}
 m_{\varphi}= \left(\frac{1}{2}\frac{\partial^2V_{\Phi}}{\partial\varphi^2}\right)^{1/2}\approx \left(\frac{M_{11}}{M_{pl}}\right)^{\alpha_{1}} 
    \left(\frac{R }{l_{pl}}\right)^{\alpha_{2}} M_{pl},
   \eq
where $\alpha_{1} = 3$ and $\alpha_{2} = 2$. For $M \simeq 1$ TeV and $R \simeq 10^{-22} \; m$, 
we find that $ m_{\varphi} \simeq   0.1\; GeV$. In the framework of string, the radion mass is also given by Eq. (\ref{2.1}) but now with
$\alpha_{1} = 8/3$ and $\alpha_{2} = 5/3$ \cite{WS08}.

 {\em Localization of Gravity and 4D Effective Newtonian Potential}:  In contrast to the RS1 model  in which the gravity is 
localized on the invisible  brane \cite{RS1},  we find that the gravity in the framework of both string \cite{WS08} and M theory
\cite{WGW08}  is localized on the visible brane, because in the present setup the warped factor increases  as one approaches the 
visible brane from the invisible one.  The spectrum of the gravitational  KK modes is discrete and the corresponding masses are 
given  by \cite{WGW08,WS08,Devin09}
\begin{equation}
m_n\simeq n\pi \left( \frac{l_{pl}}{y_c}\right) M_{pl},  \label{7.23}
\end{equation}
where $y_{c}$ is the distance between the two branes. Thus, for $y_{c} \simeq 10^{-19}\; m$ we have $m_1  \simeq 1$ TeV.

The  4D Newtonian potential, on the other hand, takes the form,
\bq
\lb{2.2}
U(r) = G_{4}\frac{M_{1}M_{2}}{r} \left(1 + \frac{M_{pl}^{2} }{M^{3}_{5}} 
       \sum^{\infty}_{n =1}{e^{-m_{n}r} \left|\psi_{n}(z_{c})\right|^{2}}\right), 
\eq
where  $\psi_{n}(y_{c}) \simeq  \sqrt{{2}/{y_{c}}}$. Clearly,  for $y_{c} \simeq 10^{-19} \; {\mbox{m}}$ and
$r \simeq 10 \; \mu{\mbox{m}}$, the high order   corrections to the 4D Newtonian potential   are exponentially  suppressed, 
and can be safely neglected. 

{\em The hierarchy problem}:  This  problem can  also be addressed in our current setups,  
but the mechanism is 
a combination of  the ADD large extra dimension \cite{ADD98} and the RS warped factor mechanisms \cite{RS1},
together with  the brane tension coupling  scenario \cite{Cline99}, and the 4D Newtonian constant is
given by \cite{WGW08,WS08},
\bq
\lb{2.3}
G_{N} =  \frac{g_{k}}{48\pi M^{\beta_{1}}R^{\beta_{2}}},
\eq
where $g_{k}$ denotes the tension of the brane,  $(\beta_{1}, \beta_{2}) = (18, 12)$ for the HW heterotic M-Theory \cite{WGW08}, and  
$(\beta_{1}, \beta_{2}) = (16, 10)$ for the string theory \cite{WS08}. 

It is interesting to note that the   4D  effective cosmological constant  can be cast in the form \cite{GWW07,WS08},  
\bq
\lb{2.4}
\rho_{\Lambda}
= \frac{\Lambda_{4}}{8\pi G_{4}}
= 3\left(\frac{M}{M_{pl}}\right)^{\beta_{1}} \left(\frac{R}{l_{pl}}\right)^{\beta_{2}}
M_{pl}^{4}.
\eq
For $R \simeq 10^{-22} \; m$ and $M_{10} \simeq 1\; TeV$, in both cases we have $\rho_{\Lambda} \sim \rho_{\Lambda, ob} 
\simeq 10^{-47}\;  GeV^{4}$.  

\begin{figure}
\centering
\includegraphics[width=8cm]{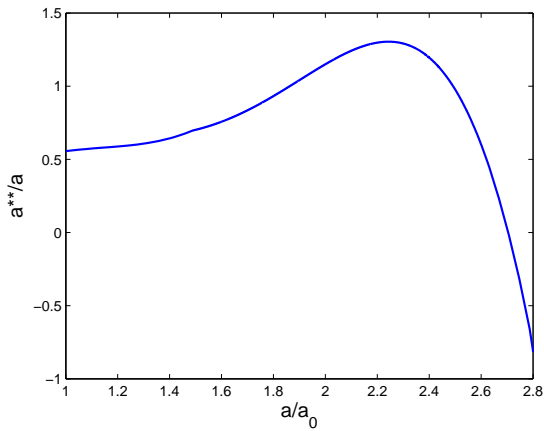}
\caption{The acceleration $a^{**}/a$  in the framework of string theory on $S^{1}/Z_{2}$ \cite{WSVW08}. }
\label{fig2}
\end{figure}

{\em Cosmological Applications}: Applying such  setups to 
cosmology, we found the generalized Friedmann-like equations on each of the two orbifold branes, and showed that the late  
acceleration of the universe is transient, due to the interaction of the bulk and the branes. 
Fig. \ref{fig2} shows the future evolution of the acceleration of the universe in the framework of string \cite{WSVW08}. 
Similar result was also obtained in M-Theory \cite{GWW07}.

Bouncing universe can be also constructed. In particular, in the  setup of M-Theory \cite{GWW07,WGW08}, the generalized Friedmann
equations for moving branes in a five-dimensional bulk with a 4-dimensional Poincare symmetry take the forms,
\bqn
\label{2.5a}
H^2 =\frac{2\pi G}{3 \rho_\Lambda}\left(\rho +\tau _\phi +2\rho _\Lambda\right)^2 - \frac {1}{25L^2a^{12}} ,\\
\label{2.5b}
\dot{\rho}+\dot{\tau_\phi}+3H(\rho+p)=-6H(2\rho_\Lambda+\rho+\tau_\phi),
\eqn
where $L$ is a constant, and $ \tau _\phi  \equiv 6\epsilon \alpha \kappa _5^{-2}e^{-\phi }
+ V_{4}(\phi)$, with $V_{4}(\phi)$ denoting the potential on the brane.  
Clearly, at the early time the last term in the right-hand side of Eq. (\ref{2.5a}) dominates, and there always exists a non-zero
minimum $a_{i} > 0$ at which $H(a_{i}) = 0$. For $ a < a_{i} $ the motion is forbidden, whereby a bouncing universe
is resulted. Similar results can be also obtained in the setup of string theory \cite{WS07,WS08,WSVW08}.

{\em Concluding Remarks}: With all these remarkable features, it is very desirable   to investigate  other aspects of 
these models. In particular,  in our previous studies, we have 
not addressed the issue of supersymmetry.  Working with TeV scale, a distinctive feature is the possibilities of finding 
observational signals to  LHC   \cite{DHR00}.  Meanwhile, in the framework of brane cosmology, 
significant deviations come from the early universe \cite{branes}. 
 To explain  the late cosmic acceleration of the universe, various models have been proposed \cite{DEs}. While different 
models can give the same late time accelerated expansion, the growth of matter perturbation they produce usually differ 
\cite{GIW09}. Recently, the use of the growth rate of matter perturbation in addition to the expansion history of the Universe 
to differentiate dark energy models and modified gravity attracted much attention. Therefore, it would be very important
to study   perturbations of the cosmological models in our string/M theory setups.    

\begin{acknowledgments}
The author would like to thank  T. Ali, R.-G. Cai, G. Cleaver, M. Devin, Y.-G. Gong, N.O. Santos, P. Vo and Q. Wu for 
collaborations.
 
\end{acknowledgments}

\end{document}